\documentclass[11pt]{article}
\usepackage
{
        amssymb,
        amsmath,
        amsthm,
        nicefrac,
}
\usepackage[letterpaper,margin=1in]{geometry}
\usepackage[unicode,colorlinks=true,citecolor=black,filecolor=black,linkcolor=black,urlcolor=black, pdfpagelabels, plainpages=false]{hyperref}

\DeclareMathOperator {\cost}  {cost}

\newcommand {\bbR}    {\mathbb{R}}

\newcommand {\calH}    {{\cal{H}}}

\newcommand{\TODO}{\textcolor{red}{TODO}}
\newcommand{\ccenter} {\operatorname{center}}

\newtheorem{theorem}{Theorem}[section]
\newtheorem{lemma}[theorem]{Lemma}

\newtheorem{corollary}[theorem]{Corollary}
\newtheorem{definition}[theorem]{Definition}

\title{Metric Perturbation Resilience}
\author{Konstantin Makarychev\\Microsoft Research \and Yury Makarychev\thanks{Supported by  NSF CAREER award CCF-1150062 and NSF award IIS-1302662.}\\TTIC}
\date{}
\begin{document}
\maketitle
\begin{abstract}
We study the notion of perturbation resilience introduced by Bilu and Linial (2010) and Awasthi, Blum, and Sheffet (2012). A clustering problem is $\alpha$-perturbation resilient if the optimal clustering does not change when we perturb all distances by a factor of at most $\alpha$. We consider a class of clustering problems with center-based objectives, which includes such problems as $k$-means, $k$-median, and $k$-center, and give an exact algorithm for clustering 2-perturbation resilient instances. Our result improves upon the result of Balcan and Liang (2016), who gave an algorithm for clustering $1+\sqrt{2}\approx2.41$ perturbation resilient instances.
Our result is tight in the sense that no polynomial-time algorithm can solve $(2-\varepsilon)$-perturbation
resilient instances unless $NP = RP$, as was shown by Balcan, Haghtalab, and White (2016).
We show that the algorithm works on instances satisfying a slightly weaker and more natural
condition than perturbation resilience, which we call metric perturbation resilience.
%%Our result holds for instances satisfying a slightly weaker and more natural condition than perturbation
%%resilience, which we call metric perturbation resilience.
\end{abstract}
\setcounter{page}{0}
\thispagestyle{empty} %%suppress the first page number (which is "0")
\pagebreak
\section{Introduction}\label{sec:intro}
In this paper, we present an exact algorithm for solving 2-perturbation resilient instances of clustering problems
with natural center-based objectives. The notion of perturbation resilience was proposed by Bilu and Linial~\cite{makarychev-BL} and Awasthi, Blum, and Sheffet~\cite{makarychev-ABS}. Informally, an instance is perturbation resilient if the optimal solution does not change when we perturb the instance. The definition was introduced in the context of beyond-the-worst-case analysis and
aims to capture a wide class of real-life instances that are computationally easier than worst-case instances.
As several authors argue, in instances arising in practice, the optimal solution is often significantly better than all
other solutions and thus does not change if we slightly perturb the instance~\cite{makarychev-BL,makarychev-bdls}.
It was shown that perturbation resilient instances of such problems as $k$-center, $k$-means, $k$-median,
clustering problems with center-based and min-sum objectives, Max Cut, Minimum Multiway Cut, and TSP
(with a sufficiently large value of the resilience parameter) can be solved exactly in polynomial-time~\cite{makarychev-ABS,makarychev-BHW2015,makarychev-BalcanLiang,makarychev-bdls,makarychev-BL,makarychev2014MMV,makarychev-MSSW};
while the worst-case instances of these problems are NP-hard.

In a clustering problem, we are given a metric space $(X,d)$ and an integer parameter $k$; our goal is
to partition $X$ into $k$ clusters $C_1,\dots, C_k$ so as to minimize the objective function $\calH(C_1, \dots, C_k; d)$
(which depends on the problem at hand).
%Perhaps, the most commonly used are the $k$-means, $k$-median, and $k$-center
%clustering problems with the following objectives:
The most well-studied and, perhaps, most interesting clustering objectives are
$k$-means, $k$-median, and $k$-center. These objectives are defined as follows. Given a clustering $C_1,\dots, C_k$,
%we find a center $c_i$ in each $C_i$ so as to minimize the following objectives:
the objective is equal to the minimum over all choices of centers $c_1\in C_1$,\dots,$c_k \in C_k$ of the following functions:
 \begin{align*}
 \calH_{means} (C_1,\dots,C_k;d) &= \sum_{i=1}^k \sum_{u\in C_i} d(u,c_i)^2;\\
 \calH_{median} (C_1,\dots,C_k;d) &= \sum_{i=1}^k \sum_{u\in C_i} d(u,c_i);\\
  \calH_{center} (C_1,\dots,C_k;d) &= \max_{i\in \{1,\dots, k\}}\max_{u\in C_i} d(u,c_i).
 \end{align*}
Note that in the optimal solution each cluster $C_i$ consists of those vertices $u$ that are closer to $c_i$ than to other
centers $c_j$; i.e. $(C_1,\dots, C_k)$ is the Voronoi partition of $X$ with centers $c_1,\dots, c_k$.
We refer to objectives satisfying this property as \textit{center-based} objectives.
We study two closely related classes of center-based objectives -- \textit{separable} center-based objectives and \textit{natural} center-based objectives (which
we discuss below and formally define in Section~\ref{sec:prelim}).
We note that $k$-means, $k$-median, and $k$-center are separable and natural center-based objectives.
\iffalse
We  discuss below two closely related classes of center-based objectives -- separable center-based objectives and natural center-based objectives;
we provide formal definitions in Section~\ref{sec:prelim}.
Roughly speaking, an objective is a separable center-based objective if it is the sum or the minimum of the costs of
individual clusters; an objective is a natural center-based objective if it only depends on distances between vertices and centers $c_i$.
\fi
%
Now we formally define perturbation resilience.
\begin{definition}\label{def:makarychev-perturbation}
Consider an instance ${\cal I} = ((X,d), \calH, k)$ of a clustering problem on a metric space $(X,d)$ with objective $\calH$.
An instance $((X,d'), \calH, k)$ is an $\alpha$-perturbation of $\cal I$ if
$$d(u,v) \leq d'(u,v) \leq \alpha d(u,v) \qquad \text{for every } u,v\in X;$$
 here, $d'$ does not have to be a metric.
An instance $\cal I$ is $\alpha$-perturbation resilient if every $\alpha$-perturbation of $\cal I$ has the same optimal clustering as $\cal I$.
We will refer to $\alpha$ as the perturbation resilience parameter.
\end{definition}
This definition does not require that the perturbation $d'$ is a metric ($d'$ does not have to satisfy the triangle inequality).
It is more natural to consider only \textit{metric} perturbations of $\cal I$ --- those perturbations in which $d'$ is a metric.
In this paper, we give the definition of metric perturbation resilience, in which we do require that $d'$ is a metric (see Definition~\ref{def:metric-stability}).
(Every $\alpha$-perturbation resilient instance is also $\alpha$-\textit{metric} perturbation resilient.)

\paragraph{Known results.}
Awasthi, Blum, and Sheffet~\cite{makarychev-ABS} initiated the study of
perturbation resilient instances of clustering problems. They offered the definition of a separable center-based objective (s.c.b.o.)
and introduced an important center proximity property (%which we discuss in this paper;
see Definition~\ref{def:proximity}).
They presented an exact algorithm for solving $3$-perturbation resilient instances of clustering problems with
s.c.b.o.; they also gave an algorithm for $(2+\sqrt{3})$-perturbation resilient instances of clustering problems with s.c.b.o.\  that have Steiner points. Additionally,
they showed that $\alpha$-perturbation resilient instances of $k$-median with Steiner points are $NP$-hard when $\alpha < 3$.
Later, Balcan and Liang~\cite{makarychev-BalcanLiang} designed an exact algorithm  for $(1+\sqrt{2})$-perturbation resilient instances of problems with s.c.b.o.,
improving the result of Awasthi, Blum, and Sheffet.
Balcan and Liang also studied clustering with the min-sum objective and $(\alpha,\varepsilon)$-perturbation resilience (a weaker notion of perturbation resilience, which we do not discuss in this paper).
Recently, Balcan, Haghtalab, and White~\cite{makarychev-BHW2015} designed an algorithm for $2$-perturbation resilient instances of symmetric and asymmetric $k$-center
and showed that there is no polynomial-time algorithm for $(2-\varepsilon)$ perturbation resilience instances unless $NP=RP$.
They also gave an algorithm for $2$-perturbation resilient instances of problems with s.c.b.o.\
satisfying a strong additional condition of cluster verifiability.
To summarize, the best known algorithm for arbitrary s.c.b.o.\ requires that the instance be $1+\sqrt{2}\approx 2.4142$ perturbation resilient~\cite{makarychev-BalcanLiang};
the best known algorithm for $k$-center requires that the instance be $2$-perturbation resilient, and this result cannot be improved~\cite{makarychev-BHW2015}.

We refer the reader to~\cite{makarychev2014MMV} for an overview of known results for stable instances of combinatorial optimization problems.

\paragraph{Our results.}
We define a class of clustering problems with \textit{natural center-based objectives}; an objective is a natural center-based objective if it is representable in the following form.
For some functions $f_c$  and $g_u(r)$
($f_c$ is a function of $c$, $g_u(r)$ is a function of $u$ and $r$; intuitively, $f_{c_i}$ is the cost of having a center at $c_i$ and $g_u(r)$ is the cost of connecting $u$ to a center at distance $r$ from $u$), we have
\begin{equation}%\label{def:nat-center-based}
\calH (C_1,\dots, C_k; d) = \min_{c_i\in C_i} \sum_{i=1}^k \Big(f_{c_i} +
\sum_{u\in C_i} g_{u}(d(u,c_i))\Big),\notag
\end{equation}
or
\begin{equation}%\label{def:nat-center-based-inf}
\calH (C_1,\dots, C_k; d) = \min_{c_i\in C_i} \max
\Big(\max_{i\in\{1,\dots, k\}} f_{c_i}
, \max_{\substack{i\in\{1,\dots, k\}\\ u\in C_i}}
g_{u}(d(u,c_i))\Big).\notag
\end{equation}
This class includes such problems as $k$-center, $k$-means, $k$-median, and metric facility location (for the first three problems: $f_c = 0$ and $g_u(r)$ equals $r^2$, $r$, and $r$ respectively).

We present a polynomial-time algorithm for $2$-metric perturbation resilient instances of clustering problems with natural center based objectives; thus, we improve the known requirement on the perturbation resilience parameter $\alpha$ from $\alpha \geq 1+\sqrt{2}\approx 2.4142$ to $\alpha \geq 2$
and relax the condition on instances from a  stronger  $\alpha$-perturbation resilience 
condition to a weaker and more natural $\alpha$-\emph{metric} perturbation resilience condition.
Our result is optimal, since
even $(2-\varepsilon)$-perturbation resilient instances of $k$-center cannot be solved in polynomial time unless $NP=RP$~\cite{makarychev-BHW2015}.
In particular, our result improves the requirement for $k$-median and $k$-means from $\alpha \geq 1+\sqrt{2}$ to $\alpha \geq 2$. 

\begin{theorem}\label{thm:main}
There exists a polynomial-time algorithm that given any $2$-metric perturbation resilient instance
$((X,d),\calH,k)$ of a clustering problem with natural center-based objective, returns the (exact) optimal
clustering of $X$.
\end{theorem}

Our algorithm is quite simple. It first runs the single-linkage algorithm to construct the minimum
spanning tree on points of $X$ and then partitions the minimum spanning tree into $k$ clusters using  dynamic programming. We note that Awasthi, Blum, and Sheffet~\cite{makarychev-ABS} also used the single-linkage algorithm
together with dynamic programming to cluster 3-perturbation resilient instances. However, their approach
is substantially different from ours: They first find a hierarchical clustering of $X$ using the single-linkage algorithm and then pick $k$ optimal clusters from this hierarchical clustering.
This approach fails for $\alpha$-perturbation resilient instances with $\alpha<3$ (see~\cite{makarychev-ABS}).
That is why, we do not use the single-linkage \emph{hierarchical} clustering in our algorithm, and, instead, partition the minimum spanning tree.

We note that the definitions of \textit{separable} and \textit{natural} center-based objectives are different. However, in Section~\ref{sec:labeling} we consider a slightly
strengthened definition of s.c.b.o.\ and show that every s.c.b.o., under this new definition, is also a natural center-based objective;
thus, our result applies to it. We are not aware of any non-pathological objective that satisfies the definition of s.c.b.o.\  but is not a natural center-based objective.

Finally, we consider clustering with s.c.b.o.\ and show that the optimal solution for every $\alpha$-metric perturbation resilient instance satisfies the $\alpha$-center proximity property; previously,
that was only known for $\alpha$-perturbation resilient instances~\cite{makarychev-ABS}. Our result implies that the algorithms by Balcan and Liang~\cite{makarychev-BalcanLiang}
and Balcan, Haghtalab, and White~\cite{makarychev-BHW2015} for clustering with s.c.b.o.\  and $k$-center, respectively, apply not only to $\alpha$-perturbation resilient but also
$\alpha$-metric perturbation resilient instances.

\paragraph{Overview.} In Section~\ref{sec:prelim}, we formally define key notions used in this paper. Then, in Section~\ref{sec:proximity}, we prove that the optimal solution for every $\alpha$-metric perturbation resilient instance satisfies the $\alpha$-center proximity property.
We use this result later in the analysis of our algorithm; also, as we noted above, it is of independent interest and implies that previously known algorithms
from~\cite{makarychev-BHW2015,{makarychev-BalcanLiang}} work under the metric permutation resilience assumption.
In Section~\ref{sec:algorithm}, we present our algorithm for solving $2$-perturbation resilient instances of problems with natural center-based objectives.
The algorithm consists of two steps. In the first step, we find a minimum spanning tree in the metric space $(X, d)$. In the second step, we partition
the tree into $k$ subtrees using dynamic programming and then output the corresponding clustering of $X$.
Finally, in Section~\ref{sec:labeling}, we prove that if a s.c.b.o. satisfies some additional properties, then it is a natural center-based objective. 
\section{Preliminaries}\label{sec:prelim}
In this section, we formally define key notions used in this paper: a clustering problem, $\alpha$-metric perturbation resilience, separable center-based and natural center-based objectives.
\begin{definition}
 An instance of a clustering problem is the tuple $((X,d), \calH, k)$ of a metric space $(X,d)$,  objective function $\calH$, and integer number $k>1$. The objective %function
 $\calH$ is  a function that given a partition of $X$ into $k$ sets $C_1,\dots, C_k$ and a metric $d$ on $X$
 returns a nonnegative real number, which we call the cost of the partition.
 \end{definition}
\noindent Given an instance of a clustering problem $((X,d), \calH, k)$, our goal is to partition
 $X$ into disjoint sets $C_1,\dots,C_k$, so as to minimize $\calH (C_1,\dots, C_k; d)$.
Now we define center-based and separable center-based objectives. We note that the definitions of
Awasthi et al.~\cite{makarychev-ABS} make several implicit assumptions that we make explicit here.
\begin{definition}\label{def:center-based}
 We say that $\calH$ is a center-based objective function %for a ground set $X$ and integer $k > 1$
 if the following properties hold.
\begin{enumerate}
\item Given a subset $S\subset X$ and distance $d_S$ on $S$, we can find the optimal center $c\in S$
      for $S$, or, if there is more than one choice of an optimal center, a set  of optimal centers $\ccenter(S, d_S)\subset S$.
(In the former case, $\operatorname{center}(S, d_S) = \{c\}$).
\item For every metric $d$ on $X$, there exists an optimal clustering $C_1,\dots, C_k$ of $X$ (i.e., a clustering that minimizes $\calH(C_1,\dots, C_k;d)$) such that
    for every two distinct clusters $C_i$, $C_j$, every point $p$ in $C_i$
    and centers $c_i \in \ccenter(C_i,d|_{C_i})$ and $c_j \in \ccenter(C_j,d|_{C_j})$
    we have $d(p,c_i)\leq d(p,c_j)$. Furthermore, if the optimal clustering $C_1,\dots, C_k$
    is unique, then $d(p,c_i) < d(p,c_j)$.

\end{enumerate}
The objective is separable if, additionally, we can define individual cluster scores so that the following holds.
\begin{itemize}
\item The cost of the clustering is either the sum (for separable sum-objectives) or maximum (for separable max-objectives) of the cluster scores.
\item The score $H(C, d|_C)$ of each cluster $C$ depends only on $C$ and $d|_C$, and can be computed in
polynomial time.
\end{itemize}
\end{definition}
In this paper, we consider a slightly narrower class of \textit{natural} center-based objectives (which we described in the introduction).
The class contains most important center-based objectives: $k$-center, $k$-means, and $k$-median, as well as the facility location objective.
We are not aware of any reasonable center-based objective that is not a natural center-based objective. Now, we formally define
natural center-based objectives.
\begin{definition}
We say that $\calH$ is a natural center-based objective function for a ground set $X$, if there
exist functions $f:X\to \bbR$ and $g:X\times \bbR\to \bbR$ such that
\begin{equation}\label{def:nat-center-based}
\calH (C_1,\dots, C_k; d) = \min_{c_i\in C_i} \sum_{i=1}^k \Big(f_{c_i} +
\sum_{u\in C_i} g_{u}(d(u,c_i))\Big),
\end{equation}
or
\begin{equation}\label{def:nat-center-based-inf}
\calH (C_1,\dots, C_k; d) = \min_{c_i\in C_i} \max
\Big(\max_{i\in\{1,\dots, k\}} f_{c_i}
, \max_{\substack{i\in\{1,\dots, k\}\\ u\in C_i}}
g_{u}(d(u,c_i))\Big).
\end{equation}
We require that the functions $f$ and $g$ be computable in polynomial time, and that %the function
$g_u$ be non-decreasing for every $u\in X$. We call points $c_i$ the centers of the clustering.
\end{definition}
%\noindent
We could have also defined an objective with an $\ell_p$-aggregate function
$\calH = (\sum_{i=1}^k \bigl(f_{c_i}^p + \sum_{u\in C_i} g_{u}(d(u,c_i))^p\bigr)^{1/p}$, but it is equivalent to
%the objective
(\ref{def:nat-center-based}) with $f'(c) = f^p(c)$ and $g'(u, r) = g^p(u, r)$.
\iffalse
In Appendix (see Lemma~\TODO), we show that any natural center-based objective satisfies
Definition~\ref{def:center-based}. Note that the $k$-median objective can be computed using
formula~(\ref{def:nat-center-based}) with $f_{c_i} = 0$ and $g_u(t) = t$; $k$-means
can be computed using~(\ref{def:nat-center-based}) with $f_{c_i} = 0$ and $g_u(t) = t^2$;
finally, $k$-center can be computed using~(\ref{def:nat-center-based-inf})
with $f_{c_i} = 0$ and $g_u(t) = t$.
\fi
%\TODO{Removed the discussion of $\lim_{q\to \infty} (\calH_q (C_1,\dots, C_k;d))^{1/q}$, not sure how we use it.}
\iffalse
 We note that any objective $\calH(C_1,\dots, C_k; d)$ of the form~(\ref{def:nat-center-based}) equals the limit
$\lim_{q\to \infty} (\calH_q (C_1,\dots, C_k;d))^{1/q}$, where
\begin{equation}
\calH_q (C_1,\dots, C_k; d) = \min_{c_i\in C_i} \sum_{i=1}^k \Big(f^q_{c_i} +
\sum_{u\in C_i} g_{u}(d(u,c_i))^q\Big).
\end{equation}
Thus every objective of the form~(\ref{def:nat-center-based}) can be replaced with an objective
of the form~(\ref{def:nat-center-based-nat}) with a sufficiently large $q$.
\fi

Now, we formally define a metric perturbation and metric perturbation resilience. Since we do not require that the objective $\calH$ is homogeneous as a function of the metric $d$,
we introduce two perturbation resilience parameters $\alpha_1$ and $\alpha_2$ in the definition, which specify by how much the distances can be contracted and expanded, respectively,
in the perturbed instances.
\begin{definition}\label{def:metric-stability}
Consider a metric space $(X, d)$. We say that a metric $d'$ is an $(\alpha_1,\alpha_2)$-metric perturbation of $(X,d)$
if $\alpha_1^{-1} d(u,v) \leq d'(u,v) \leq \alpha_2 d (u,v)$ for every $u,v\in X$.
An instance $((X, d),\calH, k)$ %of a clustering problem
 is $(\alpha_1,\alpha_2)$-metric perturbation resilient if for every $(\alpha_1, \alpha_2)$-metric perturbation $d'$ of $d$,
the unique optimal clustering for $((X,d'), \calH, k)$ is the same as for $((X,d'), \calH, k)$.

Note that the centers of clusters in the optimal solutions for  $((X,d), \calH, k)$ and $((X,d'), \calH, k)$ may differ. We
say that an instance $(\calH, (X, d), k)$ is $\alpha$-metric perturbation resilient if it is
$(\alpha,1)$-metric perturbation resilient.
\end{definition}
%\paragraph{Remark.}
%\begin{Premark}
Observe that if instance $((X,d), \calH, k)$ is $(\alpha_1,\alpha_2)$-metric perturbation resilient, then
$((X,\lambda d), \calH, k)$ is $(\lambda\alpha_1, \lambda \alpha_2)$-metric perturbation resilient for $\lambda \in [\alpha_1^{-1}, \alpha_2]$. Particularly,
if $((X, d), \calH, k)$ is $(\alpha_1,\alpha_2)$-metric perturbation resilient, then
$((X, \alpha_2 d),\calH, k)$ is $(\alpha,1)$-metric perturbation resilient and
the optimal solution for $((X, \alpha_2 d),\calH, k)$ is the same as for $((X, d), \calH, k)$.
Thus, to solve an $(\alpha_1,\alpha_2)$-metric perturbation resilient instance $((X,\lambda d), \calH, k)$,
it suffices to solve $\alpha = (\alpha_1\alpha_2)$ metric perturbation resilient instance $((X, \alpha_2 d),\calH, k)$.
Consequently, we will only consider $\alpha$-metric perturbation resilient instances in this paper.
%\end{remark}

Finally, we recall the definition of the $\alpha$-center proximity property, introduced in~\cite{makarychev-ABS}.
\begin{definition}\label{def:proximity}
We say that a clustering $C_1,\dots, C_k$ of $X$ with centers $c_1,\dots, c_k$ satisfies the $\alpha$-center proximity property if for all $i\neq j$ and $p\in C_i$,
we have
$d(p,c_j) > \alpha d(p, c_i)$.
\end{definition}

\section{Center Proximity for Metric Perturbation Resilience}\label{sec:proximity}
In this section, we prove that the (unique) optimal solution to an $\alpha$-\emph{metric} perturbation resilient
clustering problem satisfies the $\alpha$-center proximity property. Our proof is similar to the proof of
Awasthi, Blum, and Sheffet, who showed that the optimal solution to a (non-metric) $\alpha$-perturbation resilient
clustering problem satisfies the $\alpha$-center proximity property.
\begin{theorem}\label{lem:alpha-prox}
Consider an $\alpha$-metric perturbation resilient clustering problem $((X,d),\calH, k)$ with a center-based objective. Let $C_1,\dots, C_k$ be the unique optimal solution; and let $c_1,\dots, c_k$ be a set of centers of $C_1,\dots, C_k$ (that is, each $c_i\in \ccenter (C_i, d|_{C_i}))$. Then, the following $\alpha$-center proximity property holds: for all $i\neq j$ and $p\in C_i$,
we have
$d(p,c_j) > \alpha d(p, c_i)$.
\end{theorem}
\begin{proof}
Suppose that $d(p,c_j) \leq \alpha d(p, c_i)$. Let $r^* = d(p, c_i)$. Define a new metric $d'$ as follows.
Consider the complete graph on $X$. Assign length $len(u,v) = d(u,v)$ to each edge $(u,v)$ other than $(p,c_j)$.
Assign length $len(p,c_j) = r^*$ to the edge $(p,c_j)$.
Let metric $d'(u,v)$ be the shortest path metric on the complete graph on $X$ with
edge lengths $len(u,v)$.
Note that 
$d(p, c_j) \geq d(p,c_i) = r^*$ since 
$p\in C_i$ and $C_1,\dots, C_k$ is an optimal clustering.
Hence, for every $(u,v)$: $len(u,v) \leq d(u,v)$ and $d'(u,v) \leq d(u,v)$.
It is easy to see that
$$d'(u,v) = \min (d(u,v), d(u, p) + r^* + d(c_j, v), d(v, p) + r^* + d(c_j, u)).$$
Observe that since
the ratio $d(u,v)/len(u,v)$ is at most $d(p,c_j)/r^* \leq \alpha$ for all edges $(u,v)$, we have
$d(u,v)/d'(u,v) \leq \alpha$ for all $u$ and $v$. Hence, $d(u,v)\leq \alpha d'(u,v) \leq \alpha d(u,v)$, and
consequently, $d'$ is an $(\alpha,1)$-metric perturbation of~$d$.

We now show that $d'$ is equal to $d$ within the cluster $C_i$ and within the cluster $C_j$.

\begin{lemma}\label{lem:restrict-to-Ci-Cj}
For all $u,v\in C_i$, we have $d(u,v) = d'(u,v)$, and for all $u,v\in C_j$, we have $d(u,v) = d'(u,v)$.
\end{lemma}
\begin{proof}
I. Consider two points $u$, $v$ in $C_i$. We need to show that $d(u,v) = d'(u,v)$. It suffices to prove that
$$d(u,v) \leq  \min ( d(u, p) + r^* + d(c_j, v), d(v, p) + r^* + d(c_j, u)).$$
Assume without loss of generality that $d(u, p) + r^* + d(c_j, v)\leq  d(v, p) + r^* + d(c_j, u)$.
We have
$$d(u, p) + r^* + d(c_j, v) = d(u,p) + d (p, c_i) + d (c_j, v)\geq d(u, c_i) + d (c_j,v).$$
Since $v\in C_i$, we have $d(v,c_i) \leq d(v,c_j)$, and thus
$$d(u, p) + r^* + d(c_j, v) \geq d(u, c_i) + d (c_i,v) \geq d (u,v).$$

\noindent II. Consider two points $u$, $v$ in $C_j$. Similarly to the previous case, we need to show that
$d(u,v) \leq  d(u, p) + r^* + d(c_j, v)$.
Since now $u\in C_j$, we have $d(u, c_j) \leq d(u, c_i)$. Thus,
$$d(u, p) + r^* + d(c_j, v) = \big(d(u, p) + d(p, c_i)\big) + d(c_j, v) \geq  d(u, c_i) + d(c_j, v)\geq
d(u, c_j) + d(c_j, v)\geq
 d(u,v).
$$
\end{proof}

By the definition of $\alpha$-metric perturbation stability, the optimal clusterings for metrics
 $d$ and $d'$ are the same.
By Lemma~\ref{lem:restrict-to-Ci-Cj}, the distance functions $d$ and $d'$ are equal within the clusters $C_i$ and $C_j$. Hence,
the centers of $C_i$ and $C_j$ w.r.t. metric $d'$ are also points $c_i$ and $c_j$, respectively
(see Definition~\ref{def:center-based}, item 1).
Thus, by the definition of center-based objective, $d'(c_i,p) < d'(c_j,p)$, and, consequently,
$$d(c_i,p) = d'(c_i,p) < d'(c_j,p) = r^*= d (c_i,p).$$
We get a contradiction, which finishes the proof.
\end{proof}

\begin{corollary}\label{cor:closer-to-my-center}
Consider a 2-metric perturbation resilient instance. Let $C_1,\dots, C_k$ be an optimal clustering with
centers $c_1,\dots, c_k$. Then
each point $u$ in $C_i$ is closer to $c_i$ than to any point $v$ outside of $C_i$.
\end{corollary}
\begin{proof}
Suppose that $v\in C_j$ for some $j\neq i$. By the triangle inequality,
$d(u,v) \geq d(u,c_j) - d(c_j, v)$.
By Lemma~\ref{lem:alpha-prox}, $d(u,c_j)> 2 d(u,c_i)$ and $d(c_j,v) < d(c_i,v)/2$. Thus,
$$
d(u,v)\geq d(u,c_j) - d(c_j, v) > 2 d(u,c_i) - \frac{d(c_i,v)}{2} \geq
2 d(u,c_i) - \frac{d(c_i,u)+d(u,v)}{2}.
$$
In the last inequality, we used the triangle inequality $d(c_i,v) \leq d(c_i,u)+d(u,v)$.
Rearranging the terms, we get $d(u,v) > d (c_i,u)$.
\end{proof}

\section{Clustering Algorithm}\label{sec:algorithm}
In this section, we present our algorithm for solving 2-metric perturbation resilient instances of clustering problems
with natural center-based objectives. Our algorithm is based on single--linkage clustering: first, we find a minimum spanning tree on the metric space $X$ (e.g., using Kruskal's algorithm) and then run a dynamic programming algorithm on the spanning tree to find the clusters. We describe the two steps of the algorithm in Sections~\ref{sec:MST} and~\ref{sec:dynamic}.

\subsection{Minimum Spanning Tree}\label{sec:MST}
At the first phase of the algorithm, we construct a minimum spanning tree on the points of the metric space using Kruskal's algorithm. Kruskal's algorithm maintains a collection of trees. Initially, each tree is a singleton point. At every step, the algorithm finds two points closest to each other that belong to different trees and adds an edges between them. The algorithm terminates when all points belong to the same tree. We denote the obtained spanning tree by $T$.
The key observation is that each cluster $C_i$ forms a subtree of the spanning tree $T$.

\begin{lemma}\label{lem:span-tree} Each cluster $C_i$ forms a subtree of the spanning tree $T$. In other words, the unique path between every two vertices $u,v\in C_i$ does not leave the cluster $C_i$.
\end{lemma}
\begin{proof}
Let $c_i$ be the center of $C_i$. We show that the (unique) path $p$ from $u$ to $c_i$ lies in $C_i$, and, therefore, the statement of Lemma~\ref{lem:span-tree} holds. %Let $p$ be the unique path between $u$ and $c_i$ in the tree $T$.
Let $u'$ be the next vertex after $u$ on the path $p$. Consider the step at which Kruskal's algorithm added the edge $(u,u')$. At that step, $u$ and $c_i$ were in distinct connected components (as $p$ is the only path connecting $u$ and $c_i$). Thus, $d(u,u')\leq d(u,c_i)$ as otherwise the algorithm would have added the edge $(u,c_i)$ instead of $(u,u')$. By Corollary~\ref{cor:closer-to-my-center}, the inequality $d(u,u') \leq d(u,c_i)$ implies that $u'$ belongs to $C_i$. Proceeding by induction we conclude that all vertices on the path $p$ belong to~$C_i$.
\end{proof}

\subsection{Dynamic Programming Algorithm}\label{sec:dynamic}
At the second phase, we use dynamic programming to compute the optimal clustering. We root the tree $T$ at an arbitrary vertex.
We denote its root by $root$ and the subtree rooted at $u$ by $T_u$. We first assume that the tree is binary. Later, we explain how to transform any tree into a binary tree by adding dummy vertices.

The algorithm partitions the tree into (non-empty) subtrees $P_1,\dots, P_k$ and assigns a center $c_i\in P_i$ to all vertices in the subtree $P_i$ so as to minimize the objective:
\begin{equation}\label{eq:cost}
\sum_{i=1}^k f_{c_i} +  \sum_{i=1}^k\sum_{u\in P_i} g_{u} (d(u, c_i)).
\end{equation}
Lemma~\ref{lem:span-tree} implies that the optimal partitioning of $X$ is the solution to this problem.

Let $cost_u (k', c)$ be the minimum cost~(\ref{eq:cost}) of partitioning the subtree $T_u$ into $k'$ subtrees $P_1, \dots, P_k$
and choosing $k$ centers $c_1,\dots, c_k$ so that the following conditions hold:
\begin{enumerate}
% and assigning a center $c_i\in P_i$ (for $i\neq 1$) to each subtree $P_i$ subject to the constraints
\item $u\in P_1$ and $c_1 = c$ (we denote the tree that contains $u$ by $P_1$ and require that its center be $c$).
\item if $c_1\in T_u$, then $c_1\in P_1$ (if the center $c_1$ of $P_1$ lies in $T_u$, then it must be in $P_1$);
\item $c_i\in P_i$ for $i> 1$ (the center $c_i$ of every other tree $P_i$ lies in $P_i$).
\end{enumerate}
That is, we assume that $u$ belongs to the first subtree $P_1$ and that $c$ is the center for $P_1$.
Every center $c_i$ must belong to the corresponding set $P_i$ except for $c_1$.
However, if $c_1\in T_u$, then  $c_1\in P_1$.

Denote the children of vertex $u$ by $l_u$ and $r_u$ (recall that we assume that the tree is binary). The
cost $\cost_u (k', c)$ is computed using the following recursive formulas: if $c\notin T_{l_u}\cup T_{r_u}$, then
\begin{align}
\cost_u (j,c) = &f_c + g_u(d(c,u))+\min \big (\\
&\min \{\cost_{l_u} (j',c') + \cost_{r_u} (j'',c''): j'+j''=j-1,\;c'\in T_{l_u},c''\in T_{r_u}\},\label{line:lr}\\
&\min \{\cost_{l_u} (j',c') + \cost_{r_u} (j'',c)-f_c: j'+j''=j,\;c'\in T_{l_u}\},\label{line:l}\\
&\min \{\cost_{l_u} (j',c) +  \cost_{r_u} (j'',c'')-f_c: j'+j''=j,\;c''\in T_{r_u}\},\label{line:r}\\
&\min \{\cost_{l_u} (j',c) +  \cost_{r_u} (j'',c)-2f_c: j'+j''=j+1\}\label{line:no}\big).
\end{align}
\begin{quote}
If $c\in T_{l_u}$, then we remove lines (\ref{line:lr}) and (\ref{line:l}) from the formula.
If $c\in T_{r_u}$, then we remove lines (\ref{line:lr}) and (\ref{line:r}) from the formula.
\end{quote}
The first term $f_c + g_u(d(c,u))$ is the cost of opening a center in $c$ and assigning $u$ to $c$.
The lines~(\ref{line:lr}--\ref{line:no}) correspond to the following cases:\\[0.25em]
\indent (\ref{line:lr}) neither $l_u$ nor $r_u$ is in $P_1$; they are assigned to (trees $P_i$ and $P_j$ with centers) $c'$ and $c''$,\\[0.15em]
\indent (\ref{line:l}) $l_u$ is not in $P_1$, but $r_u$ is in $P_1$; they are assigned to  $c'$ and $c$,\\[0.15em]
\indent (\ref{line:r}) $l_u$ is in $P_1$, but $r_u$ is not in $P_1$; they are assigned to  $c$ and $c''$,\\[0.15em]
\indent (\ref{line:no}) both $l_u$ and $r_u$ are in $P_1$; they are assigned to $c$.\\[0.25em]
For leaves, we set $\cost_u (1,c) = f_c + g_u(d(u,c))$ and $\cost_u (j,c) = \infty$ for $j>1$.
Note that if we want to find a partitioning of $T$ into at most $k$ subtrees, we can use a slightly simpler
dynamic program. It is easy to verify that the formulas above hold.
The cost of the optimal partitioning of $T$ into $k$ subtrees equals $\min_{c\in X} \cost_{root} (k,c)$.

We now explain how to transform the tree $T$ into a binary tree. If a vertex $u$ has more than two children, we
add new dummy vertices between $u$ and its children by repeating the following procedure: take a vertex $u$ having more than two children; pick any two children of $u$: $u_1$ and $u_2$;
create a new child $v$ of $u$ and rehang subtrees $T_{u_1}$ and $T_{u_2}$ to the vertex $v$. We forbid opening
centers in dummy vertices $v$ by setting the opening cost $f_v$ to be infinity. We set the assignment costs
$g_v$ to be 0. Note that Lemma~\ref{lem:span-tree} still holds for the new tree if we place every dummy vertex in the
same part $P_i$ as its parent.

\section{Universal Separable Center-Based Clustering Objectives}
\label{sec:labeling}

%Awasthi, Blum, and Sheffet introduced the notion of separable center-based objectives.
%%\begin{definition}\label{def:separable}
%%A clustering objective is separable if we can define individual scores $H(C,d|_C)$ for each cluster
%%$C$ so that the following holds.
%%\begin{enumerate}
%%\item The cost of the clustering is either the maximum or sum of the cluster scores.
%%\item The score $H(C, d|_C)$ of each cluster $C$ depends only on $C$ and $d|_C$, and can be computed
%%in polynomial time.
%%\end{enumerate}
%%\end{definition}
In this section, we show that every separable center-based objective, satisfying some additional properties, is a
natural center-based objective. To this end, we define a \textit{universal} center-based objective and show that
every {universal} center-based objective is a natural center-based objective.

Loosely speaking, a universal center-based objective is a center-based objective that satisfies two properties,
which we will discuss now:
\begin{itemize}
\item An arbitrary center-based objective is defined on a specific set of points $X$ and can be used
to compute the cost of clustering only of the set $X$ (given a metric $d$ on $X$). \textbf{In contrast,
a \textit{universal objective} can be used to compute the cost of clustering of
\textit{any} ground set $X$.}
\item Recall that in every optimal clustering with a center-based objective each point $u\in X$ is closer to the center of its own
cluster than to the center of any other cluster. If a partition is not optimal, some points might be closer to the centers of clusters that do not contain them than to the centers of their own clusters.
Then, we can move such points to other clusters so as to minimize their distance to the cluster centers.
In fact, one of the two steps of Lloyd's algorithm does exactly this; hence, we call such a transformation
\textit{a Lloyd's improvement}.
We slightly strengthen the definition of a center-based objective by requiring that
\textbf{a \textit{universal objective} not increase when we make a Lloyd's improvement of
\emph{any} -- not necessarily optimal -- clustering}.
\end{itemize}
\iffalse
Loosely speaking, the objective is universal if we can use it to measures the cost of clustering
of any ground set $X$. The cost of the clustering should not depend on the identities of points in $X$
but may depend on the labels of points.
\fi
Now we give a few auxiliary definitions and then formally define universal center-based objectives.
Since data sets used in applications are usually labeled, we will consider ``labeled metric spaces''.
We will assume that the cost of clustering of $X$ may depend on the distances between
the points in $X$ and point labels (but not on the identities of points).

\begin{definition}[Labeled Metric Space]
A metric space labeled with a set of labels $L$ is a pair $((X,d),l)$, where $(X,d)$ is a metric space, and $l:X\to L$ is a function that assigns a label to each point in~$X$.
\end{definition}
\begin{definition}[Isomorphic Labeled Metric Spaces]
We say that two metric spaces $((X',d'), l')$ and $((X'',d''), l'')$ labeled with the same
set $L$ are isomorphic if there exists an isometry $\varphi: X'\to X''$ (i.e., $\varphi$ is
a bijection that preserves distances: $d'(u,v)= d''(\varphi(u), \varphi(v))$ for all $u,v\in X'$)
that preserves labels; i.e. $l'(u) = l''(\varphi(u))$ for all $u\in X'$.
\end{definition}
\noindent We note that in the definition above the set $L$ may be infinite. We denote the restriction of
$((X,d),l)$ to a non-empty subset $C\subset X$ by $((X,d),l)|_C$: $((X,d),l)|_C = ((C,d|_C),l|_C)$.
Note that the restriction of a metric set labeled with $L$ to a cluster $C$ is also a
metric set labeled with $L$.
\iffalse
Recall that in every optimal clustering with a center-based objective each point $u\in X$ is closer to the center of its own cluster than to the center of any other cluster (see Definition~\ref{def:center-based}). If a partition is not optimal, some points
might be closer to the centers of clusters that do not contain them than to the centers of their own clusters.
Then, we can move such points to other clusters so as to minimize their distance to the cluster centers.
In fact, one of the two steps of Lloyd's algorithm does exactly this; hence, we call such a transformation a Lloyd's improvement.
\fi
\begin{definition} Consider a clustering problem $((X,d),\calH, k)$ with a center-based objective.
We say that a clustering $C'_1,\dots, C'_k$ is a Lloyd's improvement of a clustering $C_1,\dots, C_k$ if there exists a set of centers $c_1,\dots, c_k$ of $C_1,\dots, C_k$ (i.e., each $c_i\in \ccenter(C_i,d)$) such that
\begin{itemize}
\item $c_i \in C'_i$ (a Lloyd's improvement does not move the centers to other clusters)
\item for every $x\in X$: if $x\in C_i$ and $x\in C'_j$, then $d(x,c_j)\leq d(x,c_i)$
(a Lloyd's improvement may move point $x$ from $C_i$ to $C'_j$ only if $d(c, c_j) \leq d(x, c_i)$).
\end{itemize}
\end{definition}
\noindent
%We slightly strengthen the definition of a center-based objective by requiring that the objective not increase when we
%improve \emph{any} -- not necessarily optimal -- clustering using Lloyd's algorithm.
%To formally define universal separable cluster-based clustering objectives, we first formally define labeled metric spaces and Lloyd's improvements.

\begin{definition}[Universal Objective]
We say that $\calH$ is a universal center-based clustering objective for a label set $L$, if for every
metric space $((X,d),l)$ labeled with $L$ the problem $((X,d), \calH_l,k)$ is a
clustering problem with a separable center-based objective and the following two conditions hold.
\begin{enumerate}
\item Cluster scores $H_l$ are universal (``can be used on any metric space''): Given any finite metric space $((C,d),l)$ labeled with $L$,
the function $H_l(C,d)$ returns a real number -- the cost of $C$; and
$H_{l'}(C',d') = H_{l''}(C'',d'')$ for any two isomorphic labeled metric spaces $((C',d'),l')$ and
$((C'',d''), l'')$.
\item If $C'_1,\dots, C'_k$ is a Lloyd's improvement of $C_1,\dots, C_k$, then
$\calH(C'_1,\dots, C'_k;d)\leq \calH(C_1,\dots, C_k;d)$.
\end{enumerate}
\end{definition}
Note that every %clustering problem $((X,d), \calH, k)$ with a
natural center-based objective is a universal objective. The label set
is the set of pairs $(f,g)$, where $f\in \bbR$ is a real number;
$g$ is a nondecreasing function from $\bbR^{\geq 0}$ to $\bbR$. Every point $x\in X$ is
assigned the label $l(x) = (f_x,h_x)$. The score of a cluster $C$ equals
\begin{align}
H_l(C,d) &= \min_{c\in C} \Big(f_c + \sum_{x\in C} g_x(d(x,c))\Big);\label{eq:Hl-sum}\\
H_l(C,d) &= \min_{c\in C} \Big(\max \big(f_c,  \max_{x\in C} g_x(d(x,c)\big)\Big).\label{eq:Hl-max}
\end{align}
It is easy to see that Lloyd's improvements may only decrease the cost of a clustering, since the functions $g_x$ are non-decreasing. We now show that every clustering problem with universal separable center-based objectives is
a problem with natural center-based objectives.
\begin{theorem}
I. Let $((X,d),\calH, k)$ be a clustering problem with a universal center-based separable sum-objective. Then, the scoring function
$H_l$ can be represented as (\ref{eq:Hl-sum}) for some nondecreasing functions $f$ and $g$
such that the minimum is attained when $c$ is a center of $C$; and thus
$\calH(C_1,\dots,C_k;d) = \sum_{i=1}^k H_l(C_i; d|_{C_i})$ is a natural center-based objective.

II. Let $((X,d),\calH, k)$ be a clustering problem with a universal center-based separable max-objective. If the cost
of any singleton cluster $\{x\}$ equals 0, then the scoring function $H_l$ can be represented as (\ref{eq:Hl-max})
for some nondecreasing functions $f$ and $g$
such that the minimum is attained when $c$ is a center of $C$; ; and thus
$\calH(C_1,\dots,C_k;d) = \max_{i\in\{1,\dots,k\}} H_l(C_i; d|_{C_i})$ is a natural center-based objective.
\end{theorem}
\begin{proof}
I. For every label $a\in L$ define two special labeled metric spaces $M_a$ and $M_{a,a,r}$. The metric space $M_a$ is
the metric space on a single point labeled with $a$; the metric space $M_{a,a,r}$ is the metric space on two points at distance $r$
that are both labeled with $a$. Consider a point $x\in X$ labeled with $a$.
Let $f_x = H(M_a)$ and $g_x(r) = H(M_{a,a,r})- f_x$.
We claim that the cost of any cluster $C$ equals
$$\min_{c\in C} \big(f_c + \sum_{x\in C\setminus \{c\}} g_x(d(x,c))\big),$$
and the minimum is attained when $c$ is a center of $C$. Consequently, equation (\ref{eq:Hl-sum}) holds
for functions $\tilde f_c = f_c - g_c(0)$ and $g_x$.
To prove the claim, we construct a new metric space $C_p'$ for every $p\in C$.
The metric space $C'_p$ consists of all points of $C$ and new points $x'$, which are ``duplicates'' of points $x\in C\setminus \{p\}$.
We do not change distances between points in $C$ and labels on the points from $C$. We copy the label of each $x$ to $x'$; i.e., we let $l(x')=l(x)$. We place $x'$ at the same location as the point $p$. That is, we let $d(x',y) = d(p,y)$ for all $x,y\in C$. Consider two clusterings of $C'_p$. The first clustering partitions $C_p'$ into the cluster $C$ and singleton clusters $\{x'\}$ (there is a singleton cluster for each $x\in C\setminus \{p\}$). The cost of each singleton cluster $\{x'\}$ equals $f_x$, since $\{x'\}$ is isomorphic to $M_{l(x)}$. Hence, the cost of the first clustering equals
\begin{equation}\label{eq:clust1}
H_l(C,d|_C) + \sum_{x\in C\setminus \{p\}} f_x.
\end{equation}
The second clustering partitions $C'$ into pairs of points $\{x,x'\}$ for each $x\in C\setminus \{p\}$ and the
singleton cluster $\{p\}$. The cost of each cluster $\{x,x'\}$ equals
$$H(M_{a,a,d(x,x')}) = g_x(d(x,x'))+f_x = g_x(d(x,p))+f_x,$$
since the cluster $\{x,x'\}$ is isomorphic to $M_{a,a,d(x,x')}$. The cost of $\{p\}$ equals $f_p$. Hence, the cost of the second clustering equals
\begin{equation}\label{eq:clust2}
f_p + \sum_{\substack{x\in C\setminus\{p\}}}\big( f_x + g_x (d(x,c))\big).
\end{equation}
Note that the points $x$ and $x'$ in each cluster $\{x,x'\}$ are symmetric, so we may assume that $x'$ is the center of the cluster $\{x,x'\}$. Observe that the first clustering is a Lloyd's improvement of the second clustering: To get the first clustering from the second clustering we move $x$ from each cluster $\{x,x'\}$  to the singleton
cluster $\{p\}$. We are allowed to do that since $d(x,p) = d(x,x')$. Hence, the cost of the second clustering is upper bounded by the cost of the first clustering. We have from~(\ref{eq:clust1}) and~(\ref{eq:clust2}),
\begin{equation}\label{eq:H-upper-bound}
H_l(C,d|_C) \leq  f_p + \big(\sum_{x\in C\setminus\{p\}} f_x + g_x (d(x,c))\big) -
\sum_{x\in C\setminus\{p\}} f_x= f_p + \sum_{x\in C\setminus\{p\}} g_x (d(x,c)).
\end{equation}
On the other hand, if $p$ is the center of $C$, then the second clustering is a Lloyd's improvement of
the first one. Hence, (\ref{eq:H-upper-bound}) is an equality when $p$ is the center of $C$.

It remains to show that functions $g_x$ are non-decreasing. We need to prove that for every $r_1\geq r_2\geq 0$,
$g_x(r_1)\geq g_x(r_2)$. We create a metric space on three points $x',x'',x'''$ each having the same label as $x$.
Let $d(x',x'') = r_1$, $d(x'',x''') = r_2$, $d(x',x''') = r_1+r_2$. The cost of clustering $\{x',x''\},\{x'''\}$
equals $2f_x + g_x(r_1)$; the cost of clustering $\{x'\},\{x'', x'''\}$
equals $2f_x + g_x(r_2)$. Since $r_1\geq r_2$, the second clustering is a Lloyd's improvement of the first one. Hence,
$2f_x + g_x(r_2)\leq 2f_x + g_x(r_1)$ and $g_x(r_2)\leq g_x(r_1)$. This completes the proof of part I. The proof of part II is similar.
\end{proof}

\bibliographystyle{plain}
\bibliography{stable}
\end{document}